\documentclass[a4paper,fleqn,usenatbib]{mnras}

\usepackage{mathptmx}

\usepackage[T1]{fontenc}
\usepackage{ae,aecompl}

\usepackage{graphicx}	
\usepackage{amsmath}	
\usepackage{amssymb}	

\title[FIDEOS at the ESO 1m telescope]{Precision stellar radial velocity measurements with  FIDEOS at the ESO 1-m telescope of La Silla}

\author[L. Vanzi et al.]{
L. Vanzi$^{1,2}$\thanks{E-mail: lvanzi@ing.puc.cl},
A. Zapata$^{1,2}$,
M. Flores$^{1,2}$,
R. Brahm$^{3,1}$,
M. Tala Pinto$^{4,1}$,
S. Rukdee$^{1,2}$,
\newauthor
M. Jones$^{5}$,
S. Ropert$^{1,6}$,
T. Shen$^{1}$,
S. Ramirez$^{1}$,
V. Suc$^{1,6}$,
A. Jord\'{a}n$^{7,3,8}$,
N. Espinoza$^{8,3,7}$.
\\
$^{1}$Centre of Astro-Engineering, Pontificia Universidad Catolica de Chile, Av. Vicu\~{n}a Mackenna 4860, Santiago Chile\\
$^{2}$Department of Electrical Engineering, Pontificia Universidad Catolica de Chile, Av. Vicu\~{n}a Mackenna 4860, Santiago Chile\\
$^{3}$Millennium Institute of Astrophysics, Santiago, Chile\\
$^{4}$Landessternwarte, Zentr\"{u}m f\"{u}r Astronomie der Universit\"{a}t Heidelberg, K\"{o}nigstuhl 12, D-69117, Germany\\
$^{5}$European Southern Observatory, Alonso de C\'ordova 3107, Casilla 19001, Santiago, Chile\\
$^{6}$Obstech SpA, Nueva Providencia 1881, of. 1620, Santiago Chile\\
$^{7}$Institute of Astrophysics, Pontificia Universidad Catolica de Chile, Av. Vicu\~{n}a Mackenna 4860, Santiago Chile\\
$^{8}$Max-Planck-Institut f\"{u}r Astronomie, Konigstuhl 17, Heidelberg D-69117, Germany.\\
}

\date{Accepted April 10, 2018. Received January 17, 2018}

\pubyear{2018}

\begin{document}
\label{firstpage}
\pagerange{\pageref{firstpage}--\pageref{lastpage}}
\maketitle

\begin{abstract}
We present results from the commissioning and early science programs of FIDEOS, the new high-resolution echelle spectrograph developed at the Centre of Astro Engineering of Pontificia Universidad Catolica de Chile, and recently installed at the ESO 1m telescope of La Silla. The instrument provides spectral resolution R $\sim$ 43,000 in the visible spectral range 420-800 nm, reaching a limiting magnitude of 11 in V band. Precision in the measurement of radial velocity is guaranteed by light feeding with an octagonal optical fibre, suitable mechanical isolation, thermal stabilisation, and simultaneous wavelength calibration. Currently the instrument reaches radial velocity stability  of $\sim$ 8 m/s over several consecutive nights of observation.
\end{abstract}




\section{Introduction}
High resolution spectroscopy is an extremely valuable tool in the study of a wide variety of celestial objects and phenomena. Observations of great interest can be obtained with this technique almost with any size telescope. In particular moderate size telescopes proved to be an extremely useful complement to bigger instruments when equipped with suitable spectrographs. It is mainly for this reason that our team embraced the challenge of providing the scientific community with spectrographs able to make the best use of the small and medium size telescopes available in Chile, which currently tend to be underused or decommissioned. At the same time, through this work we aim at building up an experience in astronomical instrumentation that was not present hitherto in the Chilean community.  Our first effort in this direction was the spectrograph PUCHEROS installed at the UC Observatory Santa Martina \citep{Vanzi2012}. This project proved to be a success, as it has been in operation during the last seven years, proving to be a precious tool for teaching experimental astronomy, at the undergraduate and graduated level, and also for producing scientific results, which are of surprising significance and quality when compared to the modest aperture of the telescope and to the far-from-optimal location of the observatory \citep[e.g.][]{Coronado2015, Izzo2015, Bluhm2016, Arcos2017}. This positive experience motivates our team, at the Centre of Astro Engineering UC (AIUC), to continue with energy this line of work.
In particular the measurement of high precision radial velocity (RV) is a field where relatively small aperture telescopes can provide outstanding results contributing significantly
 to a number of research fields in astrophysics as the study of binary and multiple stellar systems, astro-seismology, and the search for exoplantes. It is for these reasons that we concentrated our efforts in overcoming the limitation of PUCHEROS, mainly the quality of the site (located at about 22 km from the centre of Santiago), the small aperture of the telescope (50 cm), the low efficiency of the detector ($\sim$ 60\%), and the modest stability for RV measurements (typically 50-100 m/s). All this was casted into a new instrument called FIbre Dual Echelle Optical Spectrograph or FIDEOS, which recently saw first light at the ESO 1m telescope of La Silla. The main scientific driver of the instrument is to provide spectroscopic follow-up for photometric transit surveys searching for exoplanets around bright stars such as TESS \citep{Ricker2014}, MASCARA \citep{Talens2017}, or the upcoming HATPI\footnote{http://hatpi.org}. To achieve this purpose, FIDEOS aims at reaching a limiting magnitude V=11 and a RV precision better than 10 m/s, allowing a comfortable detection of the reflex motion induced on nearby stars by gas giant planets. One of the challenges of the project was to reach these performances within the strong constraints of the budget available for the hardware (< 100 K EUR). In this paper we present: a general description of the system in Section 2 including instrument and telescope; results from the commissioning in Section 3; results from early science programs in Section 4; and conclusions in Section 5.

\section{Instrument and Telescope}
In this section we present the general instrument concept, details of the opto-mechanical design and the general system setup. We also describe the refurbishment of the ESO 1m telescope control that was implemented by our team. The instrument includes three main parts, the spectrograph, the calibration unit, and the telescope interface. They are  connected one with another by optical fibres. To limit the cost, moderate price off-the-shelf components were selected whenever possible or were manufactured in house.
A preliminary design of FIDEOS was presented by \citet{Tala2014}, however there have been significant modifications since, which make it worth updating the instrument description here.

\subsection{Spectrograph}
The spectrograph uses a classic echelle configuration, the optical components are as follows. The collimator is a parabolic mirror of  762 mm focal length, 152.4 mm in diameter by Edmund Optics with enhanced aluminium coating and it is used off-axis . The echelle is R2.75, with 44.41 gr/mm, 110$\times$60 mm in size by Richardson gratings, it is mounted with an off axis angle $\gamma \approx 4^{\circ}$. Cross dispersion is provided by two identical prisms of SF11 glass with apex angle of $34.6^{\circ}$ manufactured by Pecchioli Research and mounted with an angle of $64.4^{\circ}$ between one an other. Finally the objective is a commercial Canon-EF-300 mm f/4 photographic objective.

The spectrograph is illuminated by two 50 $\mu$m core low-OH fibres through a tele-centric re-imaging system, which provides a magnification 4. The two fibre cross sections are octagonal to improve the scrambling effect \citep{Chazelas2010, Avila2012}, they are separated by 110 $\mu$m center-to-center and they are mounted in a single FC connector. The fibre cable was provided by Ceramoptec. One fibre brings the light from the telescope to the spectrograph (science fibre), the other carries the light from the calibration unit for simultaneous wavelength reference (calibration fiber). The telecentric system is assembled on a 30 mm ThorLabs cage stage and it includes two achromatic doublets of 7.5 and 30 mm focal length respectively, separated by the sum of their focal lengths, the beam from the fibre is received at f/5 and illuminates the collimator at f/20. The slow beam allows to obtain excellent image quality with the commercial collimator. Because the images of the fibres have a diameter of 200 $\mu$m the theoretical spectral resolution of the spectrograph is about 21000, which is boosted by a factor 2 with a 2x image slicer. Details of the slicers are presented in \citet{Tala2017}. The collimated beam is 38 mm in diameter.

The collimator and the echelle are mounted on kinematic mounts designed and manufactured at the AIUC, both supports allow tilt around two axis with micro-metric screws for precise alignment. The prism and objective are mounted on static mounts. The whole setup is arranged on a ThorLabs aluminium bread board of 900$\times$600 mm and protected by an enclosure.

As detector we employ a PL-230 Finger Lake Instruments (FLI) camera, which includes a back illuminated Midband e2v CCD $2K\times2K$ with pixel size of 15 $\mu$m. The camera comes with a convenient USB connection. Because the optics of the spectrograph has a reduction factor of 2.5, the 100 $\mu$m size half-fibre is imaged on 40 $\mu$m providing  a spectral sampling of 2.6 pixels. The FLI camera is mounted on a carbon-fibre support that guarantee stiffness and minimise the heat transfer to the optical bench and optics. This support, that was also designed and manufactured at the AIUC, allows rotation around an axis perpendicular to the optical bench and shift in the direction of the optical axis for focusing. The CCD is operated at a temperature of -35 C and read out speed of 500 kHz, in these conditions we measured a read out noise of 11 $e^-$ and a negligible dark signal, both values within the specifications of the manufacturer. The operation temperature is reached and maintained by thermoelectric cooling, the dissipated power is about 45 W and it is removed via a recirculating liquid chiller. To this effect we use a liquid chiller CW-5000 by S\&A. The chiller maintain the enclosure of the FLI camera at a temperature of 19 C with a stability of about 0.1 C peak-to-peak.

\begin{figure}
	\includegraphics[width=\columnwidth, angle=-0]{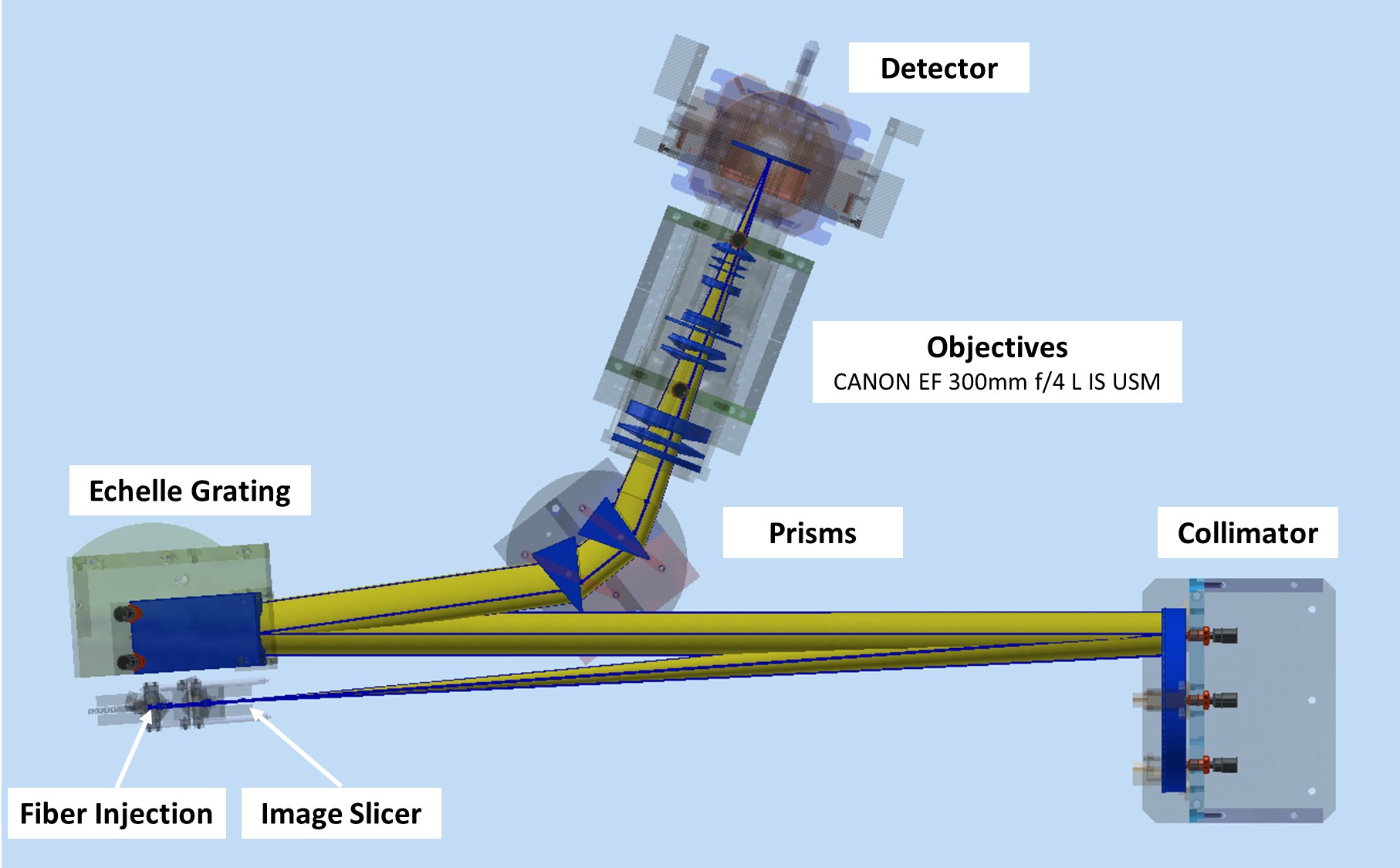}
    \caption{Top view of the opto-mechanical layout of the spectrograph, the optical components are identified by labels.  }
    \label{layout}
\end{figure}

\begin{figure}
	\includegraphics[width=\columnwidth, angle=-0]{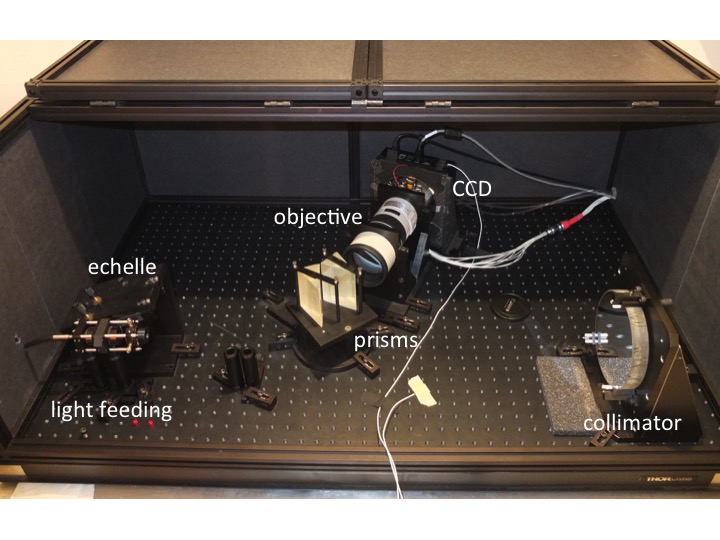}
    \caption{FIDEOS spectrograph and enclosure. In the front of the image on the left side the optical fibre with the feeding tele centric optics and the echelle are visible. To the right the collimator parabola. In the centre the two cross dispersing prisms and the objective. In the back the CCD camera.}
    \label{SPC}
\end{figure}

The top view opto-mechanical layout of the instrument, with the optical elements identified, is shown in Fig. \ref{layout}, an image of the spectrograph assembly is showed in Fig. \ref{SPC}.

\subsection{Calibration Unit}
The calibration unit includes a ThAr lamp for wavelength calibration and a continuum halogen lamp for order definition and flat fielding. The ThAr lamp is provided by Photron, the halogen lamp is a HL2000-FHSA of Ocean Optics. To improve the red to blue balance of the continuum we use a daylight blue filter LB-120 of Edmund Optics. Both lamps are imaged through the same optics on two fibres mounted on the same connector. This configuration forces to have both fibres illuminated at the same time with either lamp. One fibre goes to the spectrograph providing simultaneous wavelength reference (calibration-fibre), the other goes to the telescope interface and can illuminate the science fibre (illumination-fibre). A beam splitter is used to merge the two optical beams with a balance 90\% ThAr, 10\% halogen. The two lamps are never ON at the same time. The intensity of the ThAr lamp can be adjusted positioning a neutral density filter which is mounted on a linear stage.

\subsection{Telescope Interface}
\begin{figure}
	\includegraphics[width=\columnwidth, angle=0]{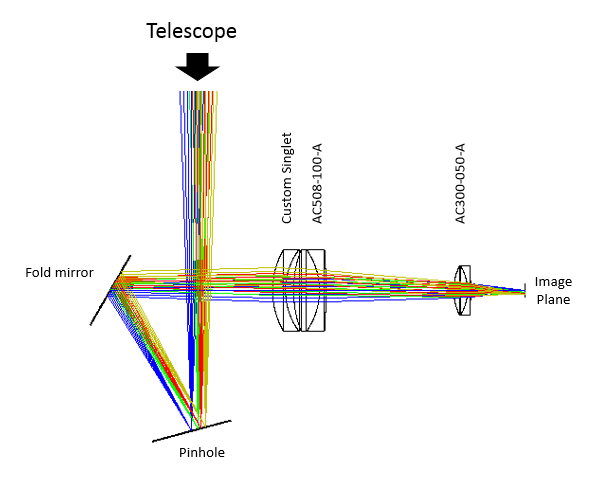}
    \caption{Detail of the acquisition optics with two folding mirrors and a split triplet. The first folding mirror has a pinhole in its centre which is the entrance of light to illuminate the science fibre.}
    \label{acquisition}
\end{figure}

The telescope interface connects  the science-fibre with the telescope making possible its illumination by the astronomical source. The light from the source enters the system trough a pinhole of 125 $\mu$m diameter, equivalent to 1.9 arcsec on the sky. The pinhole is drilled in a reflective surface which allow to re-image the field around the target on an acquisition camera to perform centring and guiding. The acquisition camera is an Imaging Source DMK33UP1300 1280$\times$1024 pix and pixels of 4.8 $\mu$m. The acquisition optics is a split triplet \citep{smith2004}, it employs a BK7 custom made lens and two commercial doublets - Thorlabs AC508-100-A and AC300-050-A. The lay out of the acquisition optics is shown in Fig. \ref{acquisition}. The effective focal length of the system is 81.3 mm and the focal working ratio is F/4.1. The field of view is 4.4$\times$3.5 arcmin with a scale of about 0.21 arcs/pix. 
The pinhole is re-imaged on the head of the science-fibre by an achromatic triplet of 12.5 mm focal length, 6.5 mm diameter by Edmund Optics. The science-fibre feeding system is mounted in a 30 mm ThorLabs cage, the science-fibre was initially supported by a five axis mount, however the tilt adjustment proved to be unnecessary and the stage was substituted with a simpler XY mount which also guarantees better stability against flexures and temperature variations .

The telescope interface includes a tip-tilt correction system for optimal light feeding. This function is provided by a commercial Starlight Express AO system, the control was developed by us. 
The centroid of the target (weighted by light intensity) is calculated from the images of the acquisition camera, the error in pixels is translated to an error in distance at the focal plane and the steps needed for the correction are calculated. The correction is applied tilting a plane parallel plate controlled in two axis by step motors. The system is controlled at a speed up to 10 Hz proving very effective for fast tracking errors and to some extent for atmospheric tip-tilt correction. The improvement in efficiency can reach up to 50\% in poor seeing conditions. 

\begin{figure}
	\includegraphics[width=\columnwidth, angle=0]{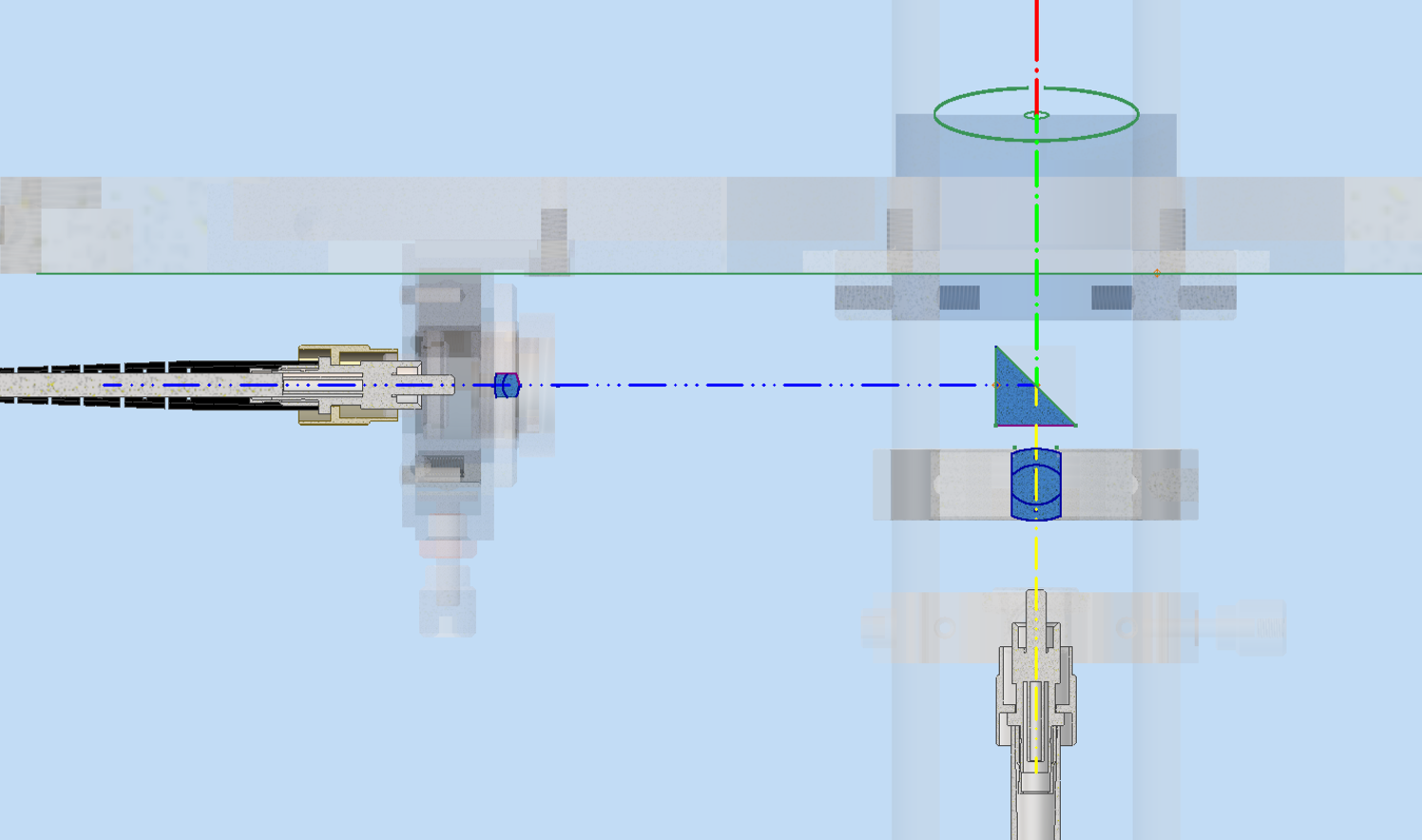}
    \caption{Detail of the fibre feeding system at the telescope interface. The beam from the telescope comes from the top of the image and reaches the science-fibre through the pinhole and the achromatic triplet. A prism can be inserted in the path (moving perpendicular to the plane of the image) to redirect the light from the illumination-fibre, on the left side of the image, to the science-fibre. When the prism is not inserted a shutter covers the illumination-fibre.}
    \label{Interface}
\end{figure}

The telescope interface also receives the illumination-fibre from the calibration unit allowing injection of the calibration light into the science-fibre. This is done with a moving prism that can be inserted between the pinhole and the fibre-feeding-optics with a linear motorised stage. Because the calibration-fibre and the illumination-fibre are always illuminated at the same time, a shutter covers the end of the illumination-fibre during the observations to avoid light contamination in the telescope interface. A sketch of the system is shown in Fig. \ref{Interface}.

The telescope interface also allows to redirect the beam of the telescope with a folding mirror toward a lateral auxiliary port that can receive a guest camera. The interface itself is an aluminium structure measuring 500$\times$250$\times$250 mm and weighting 32 Kg. 

\subsection{Instrument control and data acquisition}
There are a number of functions in the instrument that require suitable control, mainly the two calibration lamps and the neutral density filter in the calibration unit, the calibration prism in the telescope interface. There are no moving parts in the spectrograph. The neutral density filter and the calibration prism are both mounted on linear stages that refer to a home position defined by a mechanical limit switch. The low level control of the calibration module is provided by an Arduino Mega, working with an Ethernet shield which provides remote access. The calibration lamps are controlled through a relay board designed and built at AIUC. The Neutral Density filter linear stage is operated by a stepper motor controlled through a motor driver board based on the H-bridge 
driver Chip L298N. The Arduino hosts a web based engineering GUI which provides a control interface for the different devices. Integration with the 
spectrograph's main control system is given by HTTP methods and the information is formatted in JSON.

We implemented a classical instrument control software architecture, which comprises an observation control subsystem (OCS), an instrument control subsystem (ICS), and a detector control system (DCS). The OCS is the orchestrator of the observation, which interfaces the telescope (TCS) to acquire a target, configure the hardware devices through the ICS and take an exposure by commanding the DCS.  These processes were implemented on top of the Internet Communication Engine (ICE), a distributed programming framework \footnote{https://zeroc.com/products/ice}.
The scientific CCD is controlled by the DCS process through the USB connection. The DCS controls the exposure time, shutter, and temperature control loop. The acquisition camera is controlled by a second TCS process, which runs inside of an Odroid system, and continuously capture snapshots of the field. The image-processing algorithm is implemented on top of OpenCV \footnote{http://opencv.org/}.
The OCS hides the complexity of the other subsystems and exposes high-level information to the operators through a web browser.  Several dashboards were created in order to monitor the status of each subsystem during operation. A GUI was created to allow interactive observation, but at the same time, a batch mode could execute a list of calibrations in an unattended manner.

\subsection{Data processing}
The data processing is performed by a dedicated pipeline developed within the CERES framework \citep{Brahm2017}. The pipeline has basically the same structure of the CERES pipelines developed for other similar instruments (Coralie, FEROS, FIES). The steps automatically performed are: initial classification of the images according the data type (bias, flat, ThAr lamp, science image), standard image reduction routines, identification and tracing of the echelle orders for the science and comparison fibres, optimal extraction \citep{marsh1989} of the science frames and ThAr calibration images, determination of the wavelength solution for the ThAr calibration and simultaneous calibration spectra, determination of the instrumental drift in wavelength, using the simultaneous wavelength calibration mode, correction of the blaze function and continuum normalisation of the extracted science spectra. The pipeline also provides barycentric correction, measurement of the RV using the cross-correlation technique, measurement of the bisector span of the cross-correlation peak, quick determination of the stelar atmospheric parameters ($T_{eff}$, log(g), and [Fe/H]) and vsini of the science spectra.

One particular adaptation of the pipeline for this new instrument is related to the treatment of the wavelength calibration of the ThAr spectra from the calibration fiber. The profile of
an extracted narrow emission line is, at first order, symmetric in the case of the science fibre.This is because the images produced by the two slices of the fibre are aligned in the
cross-dispersion direction \citep{Tala2017}. This symmetry allows to model the line profile with a simple gaussian function, where
the mean is adopted as the position of the emission line which is then used to compute the wavelength solution. For the calibration spectrum instead only half fibre is used and this produces asymmetric lines in the extracted spectrum.The asymmetry gets partially blurred out by the instrumental profile, but nonetheless, modelling these lines with a gaussian can introduce systematic effects due to "pixelization".
For this reason we use the analytic expression of a collapsed half circumference of known diameter, convolved with a normalised
gaussian that represents the instrumental profile. The free parameters of this model are the central pixel position, the intensity of the half circumference, and
the width of the gaussian.

\begin{figure}
	\includegraphics[width=\columnwidth, angle=0]{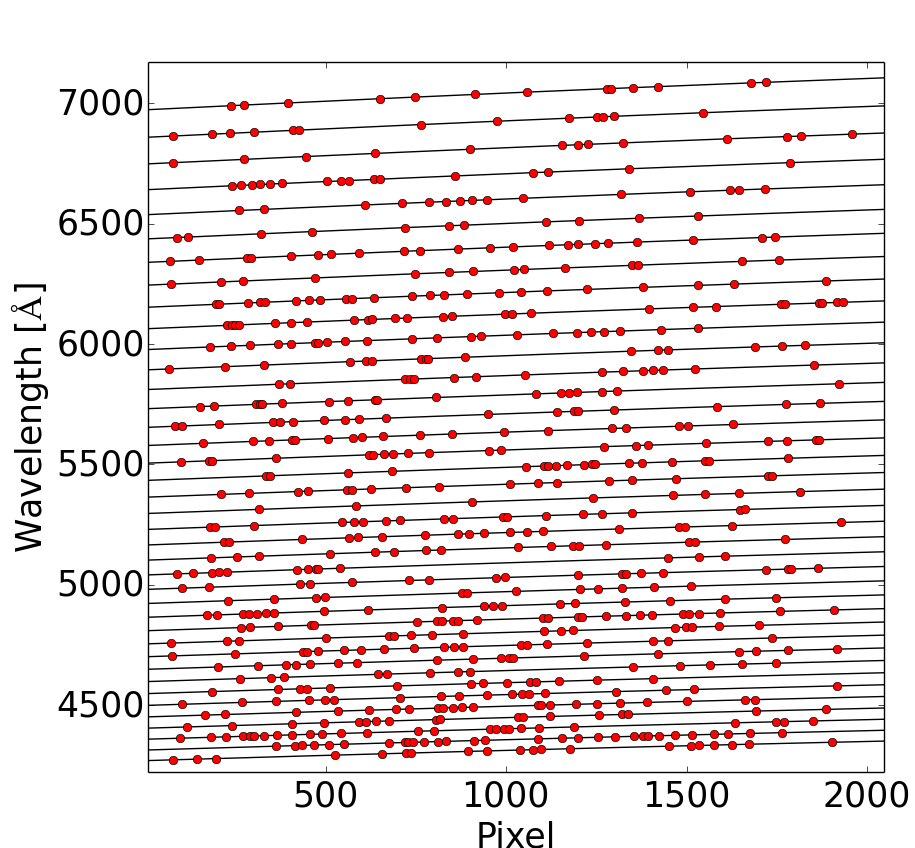}
    \caption{Representation of the global wavelength solution of FIDEOS for the different echelle orders (black line).
The red points correspond to the positions of the 690 emission lines of the ThAr lamp used to compute the wavelength solution.}
    \label{lambda-sol}
\end{figure}

\begin{figure*}
	\includegraphics[width=2\columnwidth, angle=0]{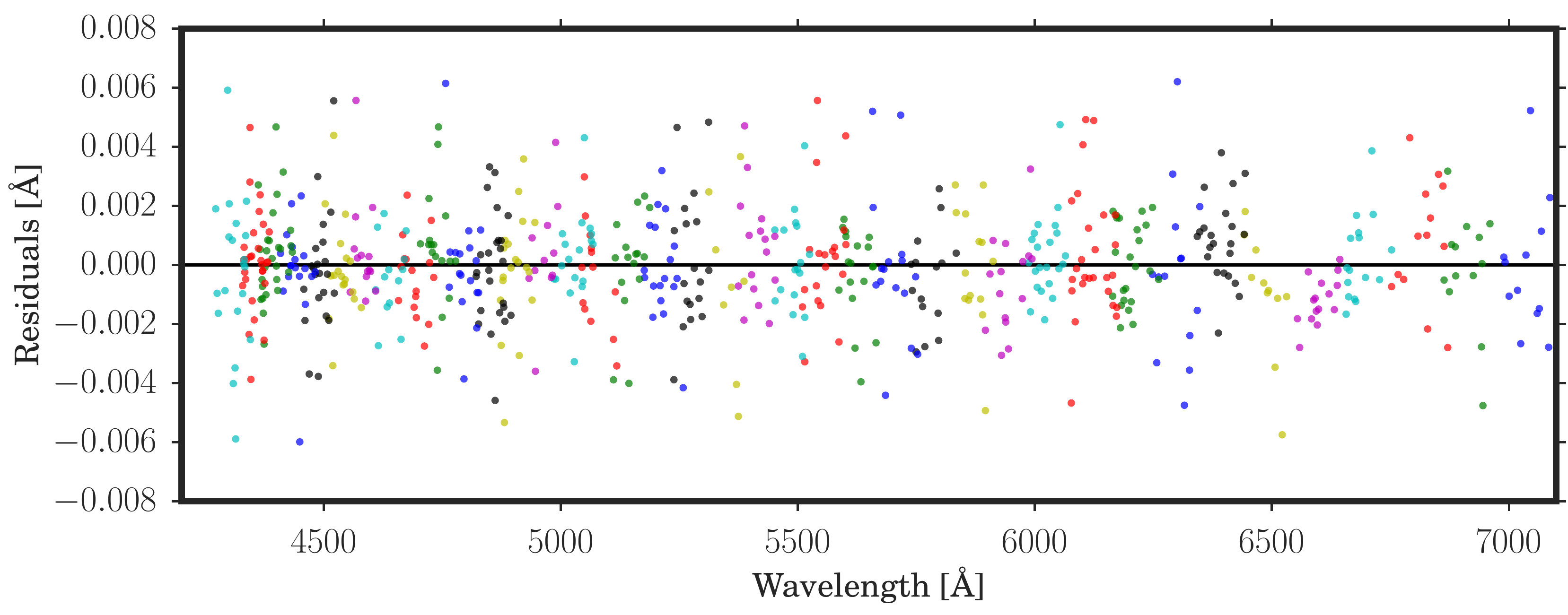}
    \caption{Residuals in $\AA$ of the wavelength solution. Each point corresponds to the difference between the reference wavelength
position of a particular ThAr line and the value given by the wavelength solution. Different orders are shown with different colours.
There is no structure left in the residuals and the typical rms is of 100 ms$^{-1}$.}
    \label{lambda-disp}
\end{figure*}

The pipeline generates two output spectra for each science frame. One contains the full wavelength coverage of the instrument considering the 50 echelle orders,
while the second output spectrum contains only the data of the last 39 orders, where the 11 reddest orders are not considered because of significant contamination from saturated ThAr lines. 
For computing the wavelength solution we use $\sim$ 690 ThAr emission lines, typically only a small fraction of them (no more than 20)
are automatically rejected from the fit when their positions are considered as outlier. The rejection of this small fraction of lines does not introduce systematic effects detectable at the level of best precision that we can reach. The typical root mean square of the residuals of the
final wavelength solution is of $\sim$ 100 m s$^{-1}$ which sets the lower limit of the precision that FIDEOS can achieve in RV to 4.0 m s$^{-1}$ (see Fig. \ref{lambda-sol}, and \ref{lambda-disp}).

\subsection{Instrument setup}
The setup of the instrument is arranged to optimise stability and efficiency.
The spectrograph and calibration unit are conveniently installed in a thermally isolated room at the first floor of the telescope building (FIDEOS room). This location, one floor above ground, one floor below the telescope level, is particularly favourable for maintaining a good thermal control of the spectrograph. The building walls are thick and provide by themselves some level of isolation this was further improved by covering the walls of the FIDEOS room with polystyrene panels. The only sources of heat present in the FIDEOS room are the scientific CCD, the calibration lamps and the neutral density filter motor of the calibration unit. The room has no windows and can only be accessed through a larger room (FIDEOS Service room), which is maintained at 16 C with a regular air conditioning system. The power supply of the scientific CCD, power supply of the calibration unit, acquisition and operation computer and the chiller are all located in the service room. The CCD chiller is operated at 14 C. The spectrograph is protected by a ThorLab black cardboard enclosure which is contained in a larger polystyrene box for optimal thermal isolation. Therefore there are two layers of thermal isolation. The optical breadboard sits on top of an aluminium plate which is actively stabilised in temperature by a Belektronig HAT control model M-20.

To achieve optimal throughput, a critical part of the instrumental setup is the alignment of the science-fibre at the telescope interface. To this purpose we proceed in two steps: i) we back illuminate the science fiber and observe its image through the pinhole with a microscope to optimise the centring, ii) we project a laser beam from the science fibre toward the M2 to optimise the tilt angles.

The cable length between the FIDEOS room and the telescope interface is about 30 m. As explained, we have three optical fibres, the science-fibre running from the telescope interface to the spectrograph (30 m), the calibration-fibre running from the calibration unit to the spectrograph (5m), and the illumination-fibre running from the calibration unit to the telescope interface (30 m) with the purpose of illuminating the science fibre.

\subsection{ESO 1m telescope}
The ESO 1m telescope is the first telescope that was installed by ESO at the observatory La Silla back in 1966. It is a classic Cassegrain design with 1 meter clear aperture, focal ratio 13.6, and it provides a scale of 15 arcsec/mm. It was used as photometric telescope until 1994, later employed in the DENIS survey from 1996 to 2001 \citep{Denis1997} and then decommissioned. Starting in 2013 the telescope is being operated by Universidad Catolica del Norte (UCN). As part of our work to equip the telescope with a new spectrograph we also installed a completely new control system which provides stable and reliable operation.

The original Telescope Control System (TCS) was running in an HP1000 computer which was obsolete and did not allow proper remote access and control. The new TCS is conformed by a modern and robust software running on a group of single board computers interacting together as a network with the CoolObs TCS developed by ObsTech \citep{ropert2016tcs}. The original motors were maintained and the only mechanical modifications performed were the upgrade of the existing motor encoders and the installation of new on-axis  encoders on both RA and DEC. The new TCS allows to combine the input signals of 2 encoders per axis, the telescope axis encoder and the motor encoder, using a Extended Kalman Filter \citep{kalman1960new, kalman1961new}. This configuration achieves a precision of 0.25 arc-second on sky during tracking with moderate cost encoders.

The initial pointing model, embedded in the CoolObs TCS, was obtained with a set of 15 known stars uniformly spread in the sky. This resulted in pointing residuals below 4.5 arc-seconds RMS, which allows us to point the star within the guiding camera field of view and very close to the fibre. Automatic re-centring of star on the fibre can then be performed by the guiding software. 

\section{Commissioning and Performances}
FIDEOS was installed at the telescope and had first light in June 2016. In this section we report on the main performances of the instrument, in particular the spectral resolution, throughput and stability.

\subsection{Spectral resolution and throughput}
\begin{figure}
	\includegraphics[width=0.8\columnwidth, angle=-90]{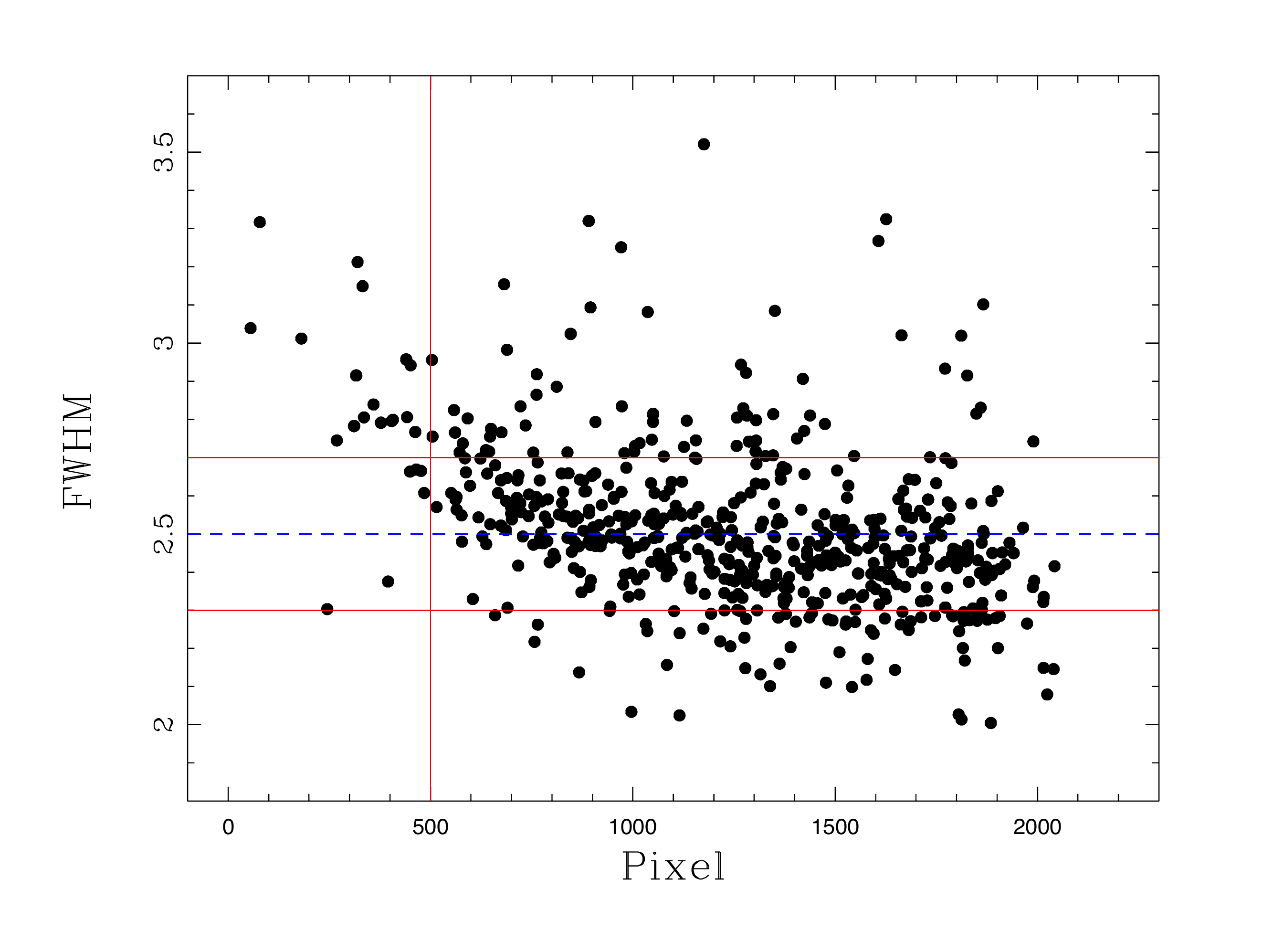}
	\includegraphics[width=1.1\columnwidth, angle=0]{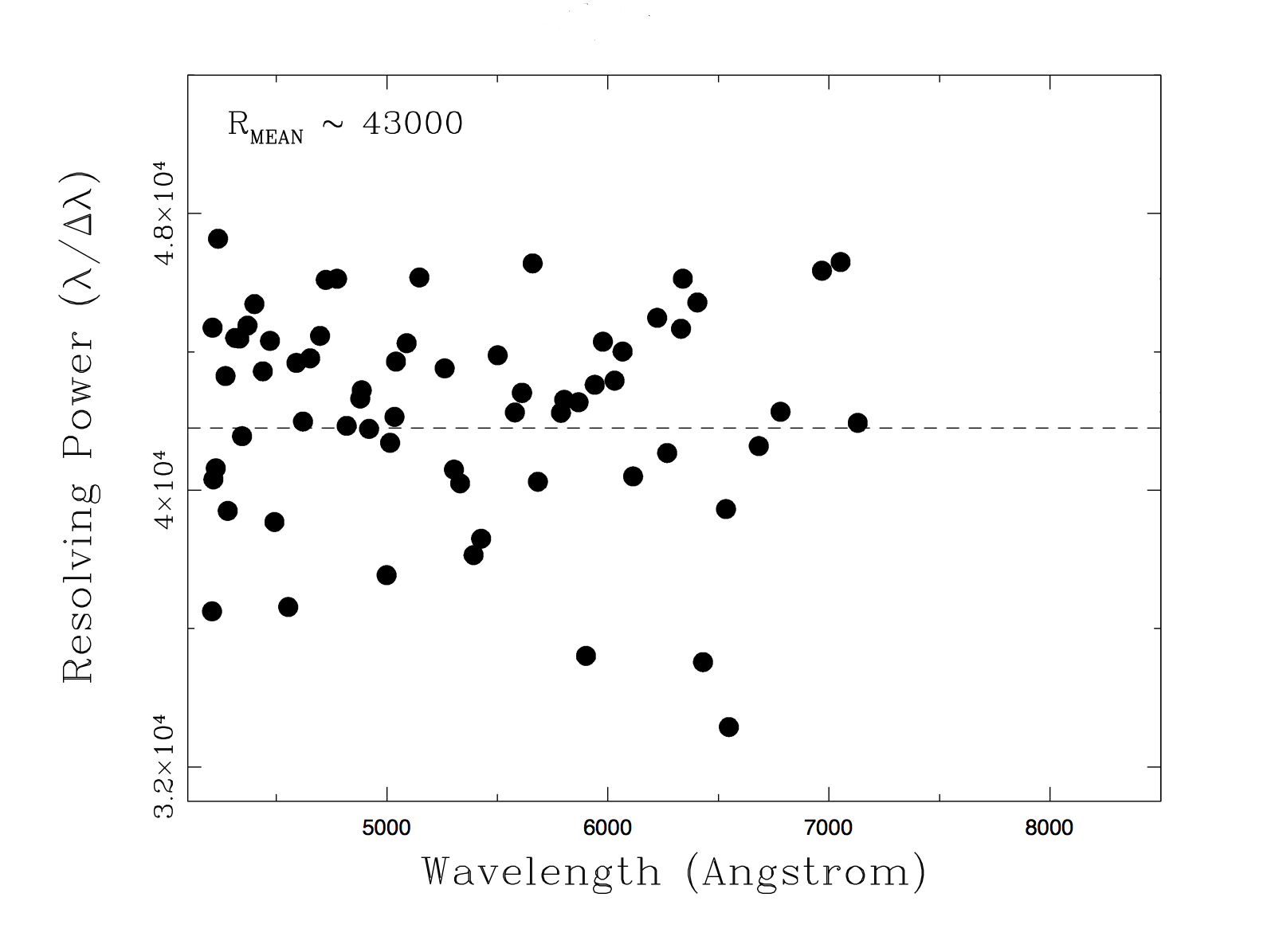}
    \caption{Full with at half maximum of ThAr emission lines across the detector versus the position in pixels on the axis of main spectral dispersion (top panel). The theoretical spectral sampling is 2.6 pixels. Spectral resolution versus wavelength (bottom panel).}
    \label{SpcRes}
\end{figure}

The spectral resolution was measured from ThAr spectra observed through the science- and calibration-fibre and measuring the FWHM for all emission lines detected, the results are shown in Fig. \ref{SpcRes}. 

During the commissioning we measured a number of spectro-photometric standard stars with different magnitudes from the catalog of \cite{Hamuy92, Hamuy94}. The results are highly dependent on the atmospheric conditions and, when the sky is clear, on the seeing. The best efficiency of the whole system turned to be approximately 8\% at 550 nm. Comparing this number with expectation is not trivial because of the large uncertainties on the efficiency of some of the components. The telescope M1 mirror gave 89\% when recently aluminised, so that we assumed an average value of 75\% for the efficiency of the telescope. The fibre link has roughly an efficiency of 60\% which is primarily driven by focal ratio degradation and, at a lower level, by absorption through the fibre and reflection at the fibre end surfaces. The maximum efficiency of the whole spectrograph, including the detector, measured in the Lab, is approximately 20\%.
The dependency on the seeing is strong. Because the aperture is almost 2 arc sec in diameter on the sky, we collect 98\% of the energy with seeing of 0.8 arcesec, 92\% with seeing 1 arcs, 82\% with 1.2 arc sec. Guiding is also a critical aspect which is largely improved with the tip-tilt system. These values provide a total efficiency of the system fully consistent with the value reported above. We also measured that the efficiency at H$\alpha$ is 80\% of the peak, at H$\beta$ 70\%, while 50\% of the peak is reached at 455 and 730 nm respectively on the blue and red side of the spectrum.

\begin{figure}
	\includegraphics[width=1.1\columnwidth, angle=0]{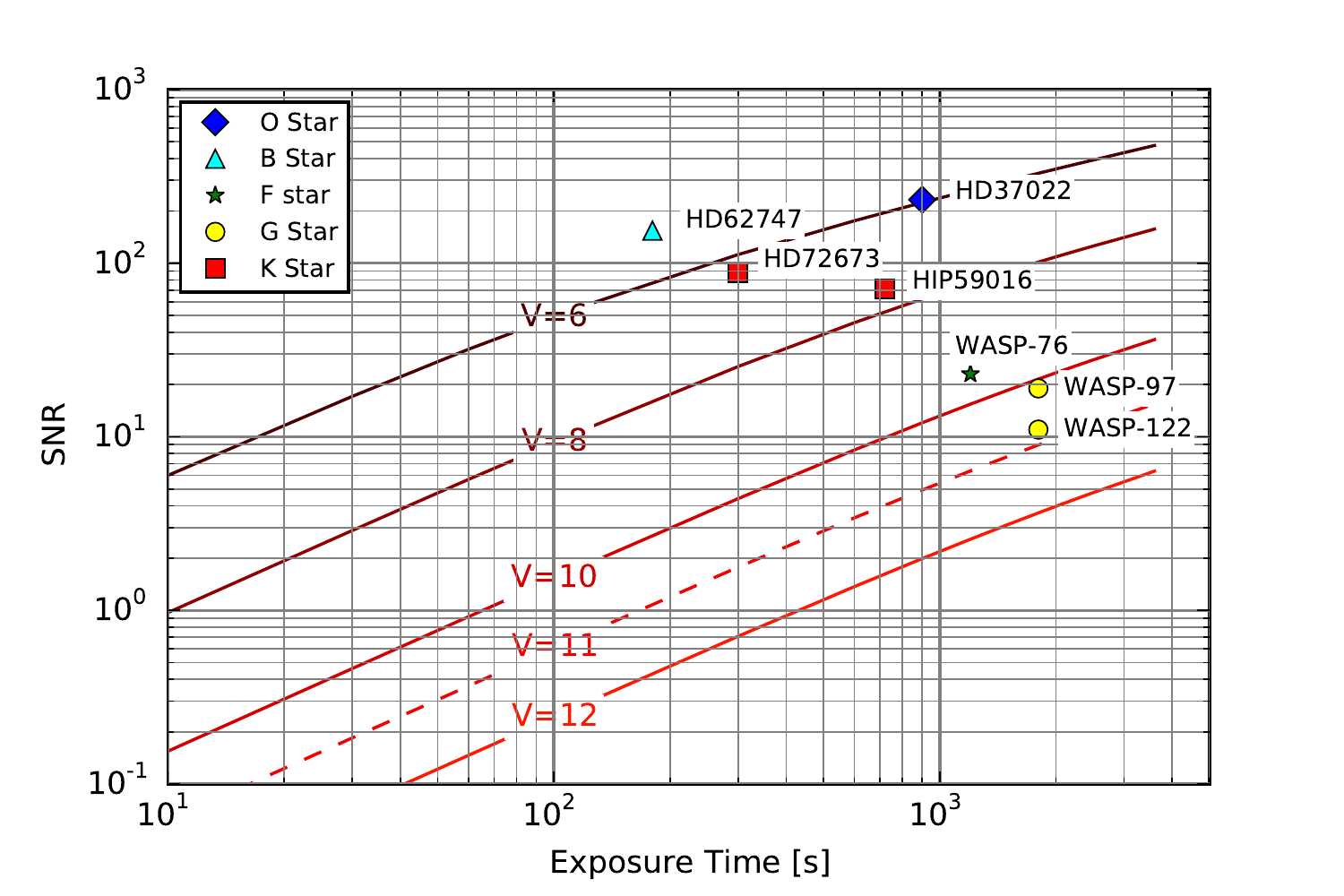}
    \caption{SNR versus integration time for V magnitudes ranging from 6 to 12. The plot is calculated for average observing conditions. The V magnitudes of the observations plotted are: HD 37022 5.13, HD62747 5.6, HD72673 6.38, Hip59016 7.04, WASP-76 9.52, WASP-97 10.57, and WASP-122 11.0.}
    \label{SN}
\end{figure}

With a read out noise of 11 $e^-$, the photon noise regime for a star of magnitude V=11, is achieved under good conditions in about 15 minutes of integration time. In Fig. \ref{SN} we show the expected signal-to-noise-ratio (SNR) versus exposure time under average observing conditions for V magnitude from 6 to 12. The theoretical expectation are compared with a few measured points. We assume as typical limiting magnitude V=11 for which SNR of about 10 can be reached in 30 minutes of exposure.


\subsection{Thermal stability}
\begin{figure}
	\includegraphics[width=1.1\columnwidth]{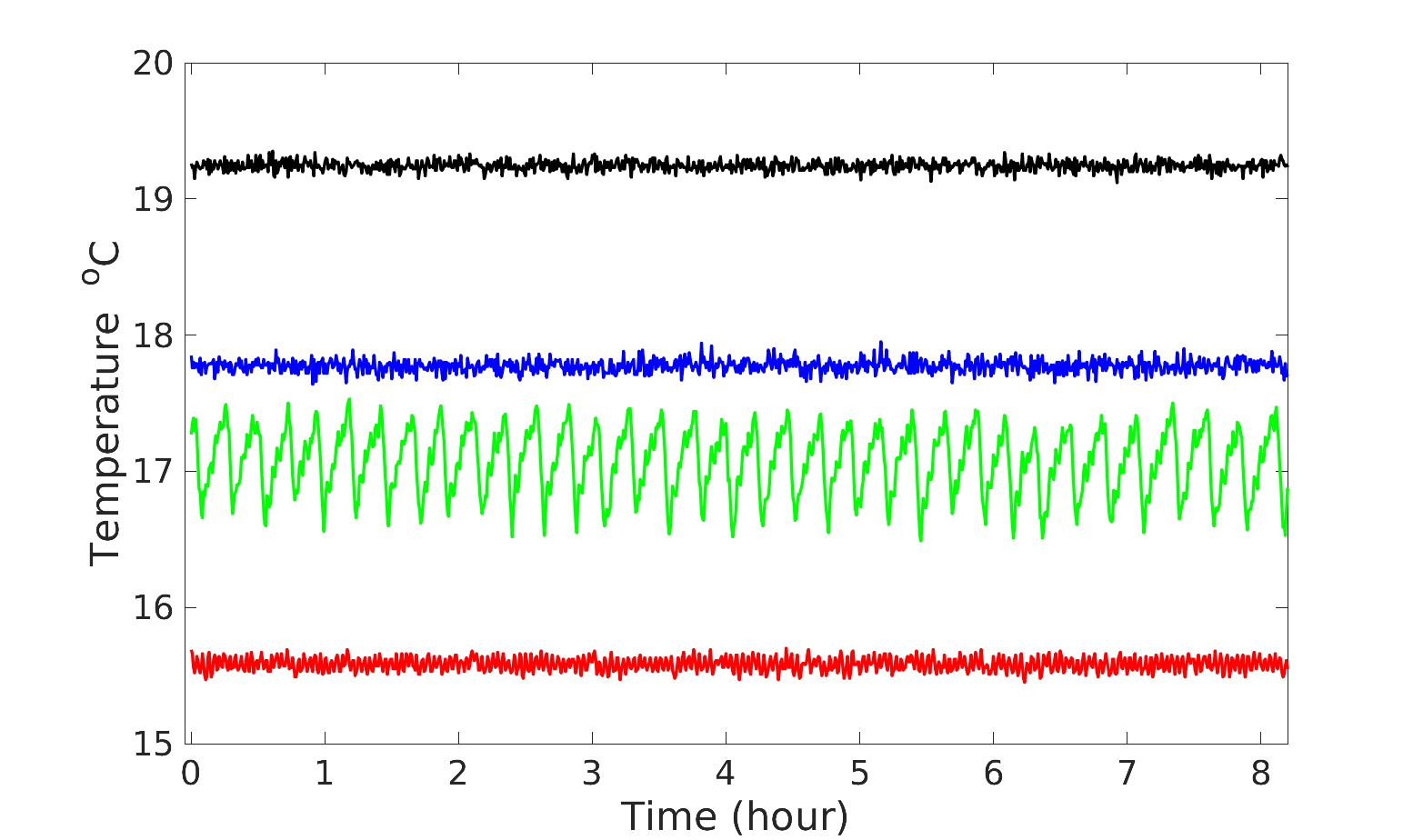}
    \caption{Temperature monitoring of the instrument versus time. The temperature is measured at four points, from the top the CCD enclosure (black), the optical bench (blue), the FIDEOS service room (green), the chiller liquid (red).}
    \label{Tmonit}
\end{figure}

The stability of the instrument is the most critical aspect when aiming at precise RV measurements, for this reason we devoted a significant amount of effort in reaching the best possible results on this front. We have been monitoring carefully the temperature at four critical points of the system, the air in the FIDEOS service room, the liquid of the recirculating chiller, the enclosure of the CCD camera, and the spectrograph optical bench. The air conditioning system is set to maintain the FIDEOS service room at a temperature of 16 C, the fluctuations around this set-point are smaller that 1 C peak to peak, in fact typically 0.6 C. The chiller is set at 14 C, the manufacturer guarantees 0.3 C stability. The HAT control is set at 17 C. The detector is operated at -35 C and stabilised within 0.1 C. Despite this promising starting point, we soon discovered that the enclosure of the FLI CCD camera was subject to strong heating, up to 1-2 C, during the exposures and that this had a substantial effect on the RV measurements, being the main source of thermal instability in the whole system. The reason for this heating was the mono-stable shutter which came by default with the camera and that is constantly powered during the exposures to stay open. As a consequence the power injected into the system by this effect is highly variable depending on the observing activity. The effect on RV turned to be of several 100 m/s, indeed similar to what can be observed with PUCHEROS \citep{Vanzi2012}. This strong effect can be attributed to a shift of the CCD on the focal plane produced by thermal expansion, in fact the detector is mechanically supported through its enclosure. To control this problem we removed the shutter from the main body of the camera and replace it with a bistable shutter from Uniblitz installed at the exit of the image slicer. This action removed completely the undesired effect. In Fig. \ref{Tmonit} we show the temperature monitored on the four points mentioned above over a period of 8 hours. The Fourier analysis of these data allows to detect interesting trends. We find that the Service Room has a principal cycle of about 900 sec, which is the main cycle of the air conditioning, and a secondary cycle of 150 sec, however the precise duration of the cycle depends on the external conditions. The chiller is dominated by a cycle of about 150 sec which is clearly transmitted to the temperature of the Service room, the fluctuations of the chiller liquid at the measuring point are below 0.15 C. The CCD enclosure presents both cycles but at quite low level, below 0.1 C which in fact is the detection limit of the temperature sensor. Finally the optical bench does not present any cycle and only shows the 0.1 C noise of the sensor.

\begin{figure}
	\includegraphics[width=\columnwidth]{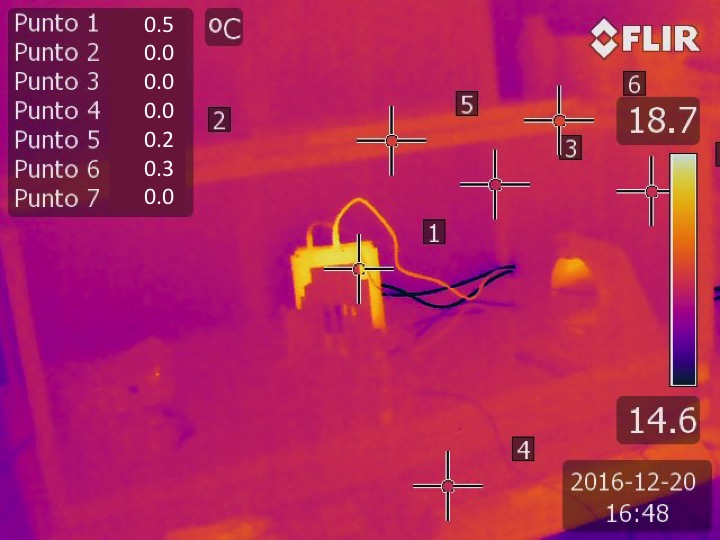}
    \caption{Thermal image of the whole spectrograph and enclosure. The CCD camera is visible as the brightest (warmest) square object in the image. The dark cables are the liquid chiller hoses, the bright cables are the power supply and USB cable. The camera is partially covered by the objective. On the right the collimator, to the left the fibre light injecting system and the echelle. All these elements are barely visible because well thermalised with the environment. The temperature differences at selected points are indicated. It is convenient to compare this image with Fig. \ref{SPC}.}
    \label{ThermalFideos}
\end{figure}

\begin{figure}
	\includegraphics[width=\columnwidth]{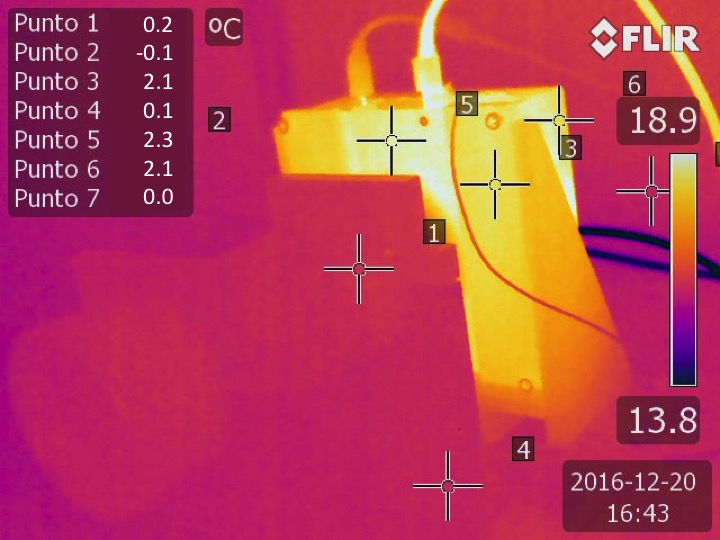}
    \caption{Thermal image of the CCD camera and objective.}
    \label{ThermalCCD}
\end{figure}

Despite the fact that we measure the temperature of the spectrograph only at two points, we made an extensive study of the instrument with a thermal camera. In Fig. \ref{ThermalFideos} a thermal image of the spectrograph with the enclosure opened is shown, the temperature is uniform across the spectrograph within 0.2 C peak to peak and it is expected to be better when the enclosure and the room are closed and stabilised. The only "hot" part is the CCD camera. In Fig. \ref{ThermalCCD} a zoom of the camera and objective can be seen.

\subsection{RV Precision}
\begin{figure*}
	\includegraphics[width=2\columnwidth]{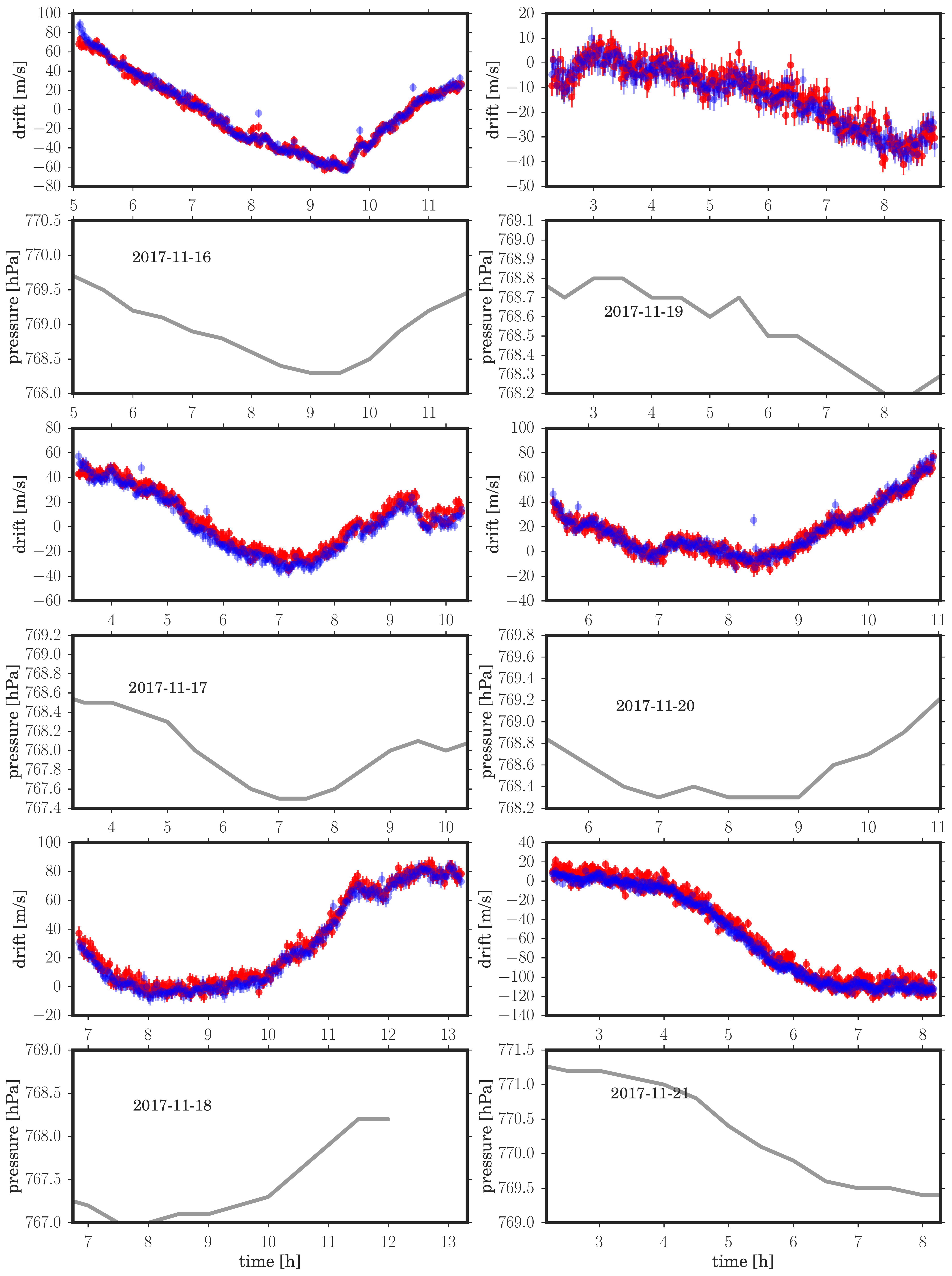}
    \caption{Velocity drift in the spectrograph observed with ThAr spectra obtained through the science (red) and calibration fibre (blue) during 6 different nights. The continuous line represents the atmospheric pressure as measured by the ESO La Silla metro station during the tests. The calibration fiber has higher SNR because it is illuminated through a much shorter optical path and receives a higher signal.}
    \label{RV-P}
\end{figure*}

With the current configuration, FIDEOS can be stabilised in temperature with a precision better than 0.1 C, however
instrumental velocity drifts can still be produced by a number of other effects. The simultaneous calibration technique
has the objective of tracing the instrumental
drift. In order to confirm that the comparison fiber is able to follow this possible drifts properly, we run a simple
experiment where we obtained a long sequence of successive calibration spectra with both fibres illuminated by the ThAr lamp.
The experiment was run during night time to approach as much as possible the real observing conditions. We repeated the
experiment for 6 consecutive nights. Figure \ref{RV-P} shows the instrumental drift measured for the science and comparison fibres for
the six different nights. There are three important findings to highlight.

\begin{figure}
	\includegraphics[width=\columnwidth]{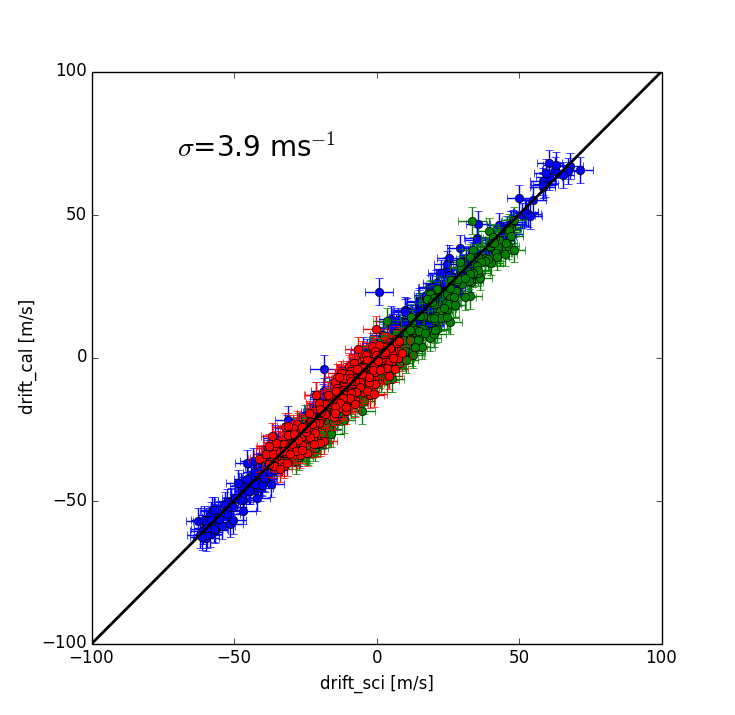}
    \caption{Velocity drift in the spectrograph observed with ThAr spectra obtained through the science (horizontal axis) and calibration fibre (vertical axis).}
    \label{RV-sigma}
\end{figure}

\begin{itemize}
\item We find a strong correlation between the instrumental velocity drift and the atmospheric pressure at ground level, measured by
the ESO La Silla weather station. Figure \ref{RV-P} proves that the ambient pressure dominates the current instrumental stability. 
Typically, a variation of 1 hPa in pressure translates in a velocity drift of 100 ms$^{-1}$, which corresponds to 3.7\% of
a pixel.
\item The variations produced by changes in pressure during our experiment were relatively smooth. Under
these conditions the use of the comparison fiber can be trusted as a tracer of the instrumental stability
for measuring precise RVs. Figure \ref{RV-sigma} shows the drift produced on the calibration fiber as
a function of the corresponding drift in the science fiber, where a strong 1:1 correlation can be identified.
We obtained that this correlation presents an rms of 3.9 ms$^{-1}$, which is fully consistent with the error obtained
for the wavelength solution, proving again the RV limit of $\sim$4 ms$^{-1}$.
\item For the two first nights of our experiment we started the acquisition of the spectra immediately after switching on the ThAr lamp.
For these nights we can identify a systematic offset for the instrumental drift computed with both fibres in the case of the
first $\sim$10 images. This shows that for achieving the $\sim$4 ms$^{-1}$ stability in RV it is necessary to wait for the
stabilisation of the lamp before starting the observations. The stabilisation of the ThAr lamp is reached after $\approx$600s.
\end{itemize}

\begin{figure}
	\includegraphics[width=1.0\columnwidth, angle=0]{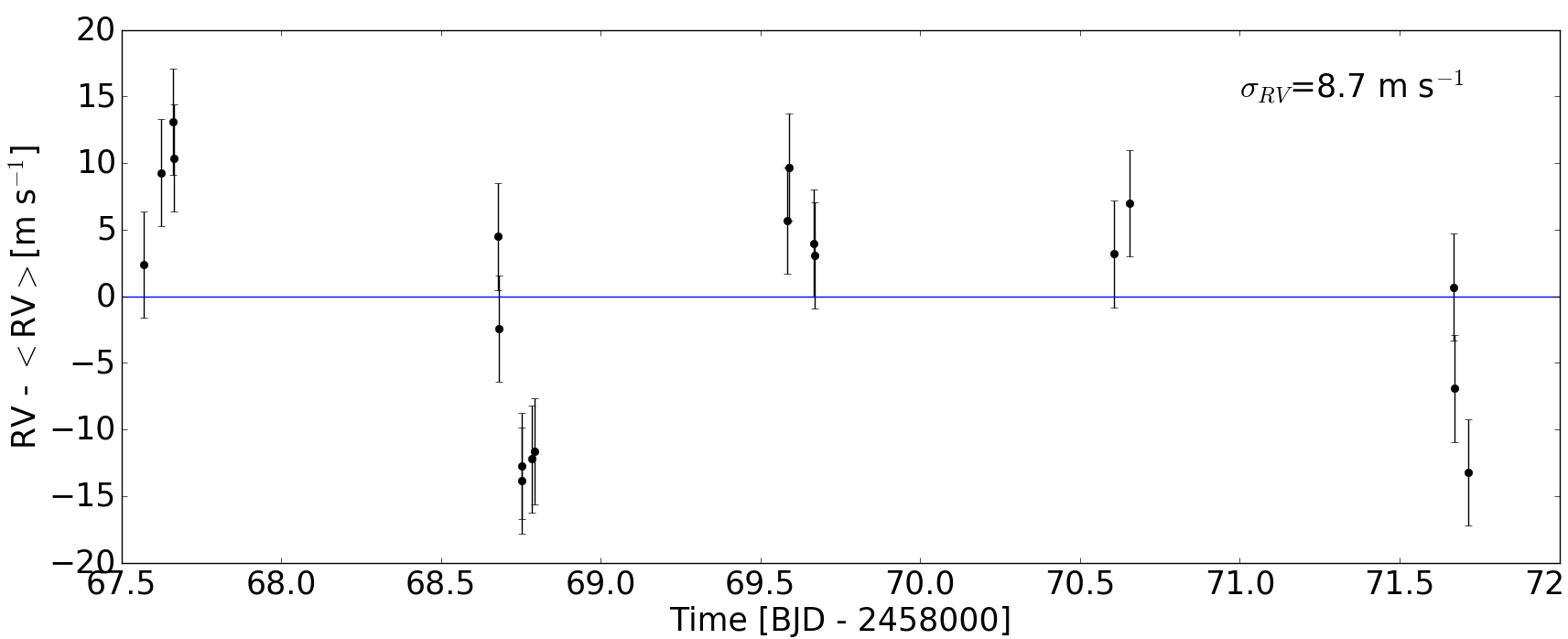}
    \caption{RV shift measured using the RV standard stars HD10700 ($\tau$ Cet) over five nights of observation. The error bars indicate $\pm$ 4 m/s.}
    \label{tauceti}
\end{figure}

\begin{figure}
	\includegraphics[width=1.0\columnwidth]{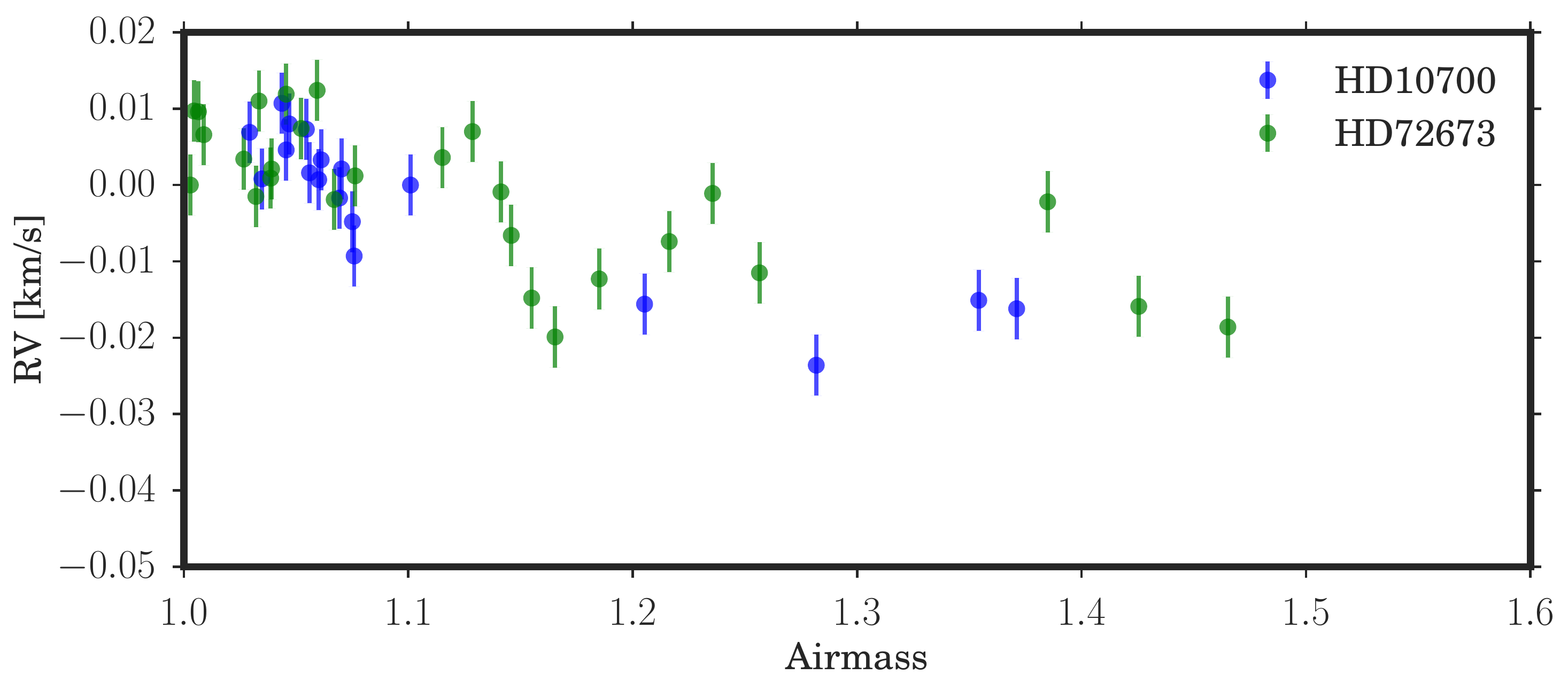}
    \caption{Measurements of RV - <RV>  versus airmass for the RV standard stars HD10700 and HD72673}
    \label{am}
\end{figure}

Additionally, we tested the stability of the system by observing the radial velocity standard star HD10700. We performed between
3 and 5 observations per night for a total number of 5 consecutive nights. We adopted an exposure time of 180s. The observations
were performed using the simultaneous calibration mode for airmass values between 1 and 1.7. The achieved SNR ranged between
150 and 340 per resolution element in the central part of the echelle order centred at 515 nm. The error on the RV expected for these values of SNR was calculated as described in \citep{Brahm2017} and it is < 2 m/s. Table \ref{tab-tau} presents
the measured RVs and bisector spans. The RV time series is presented in figure \ref{tauceti}. The RVs present a rms of  8.7 m/s.
While this precision is enough for detecting giant extrasolar planets, it is still larger that the lower stability limit
of 4.0 m/s obtained from the error in the wavelength solution and in the scatter of the instrumental drift by the
comparison fiber. The origin of the increased scatter for the measurements obtained on the sky can be produced by atmospheric dispersion, basically because the blue-to-red balance of the light entering the spectrograph is affected when observing at relatively high airmasses. We do not employ an atmospheric dispersion corrector. \cite{Bouchy2013} show that atmospheric dispersion can produce systematic offsets in RV as large as a few m/s.
To test the effect in more detail we obtained observations of the RV standard stars HD10700 and HD 72673 in a range of air masses between 1 and 2. The observations obtained with AM < 1.1 gave an RMS dispersion of 5 m/sec, close to the best expected value. However, when including all measurements the RMS was about 10 m/s. These results are shown in Fig. \ref{am}.


\begin{table}
\centering
\caption{Observations of the RV standard HD10700 ($\tau$ Cet). The columns are, BJD epoch, measured RV in km/s, Bisector span, signal-to-noise ratio and airmass}
\label{tab-tau}
\begin{tabular}{llrlll} 
\hline
BJD-2458000 &  RV  & BS  &  BSe  &  SNR  &  AM   \\
\hline
67.57031  & -16.6194  & -0.004 & 0.003  & 320  &  1.10 \\
67.62442  & -16.6135  & -0.001 & 0.003  & 324  &  1.02 \\
67.66263  & -16.6089  & -0.003 & 0.003  & 269  &  1.04 \\
67.66557  & -16.6120  &  0.000 & 0.003  & 272  &  1.04 \\
67.78355  & -16.6138  & -0.010 & 0.004  & 177  & 1.60 \\
67.81831  & -16.6270  & -0.014 & 0.005  & 148  & 2.15 \\
68.67845  & -16.6064  & -0.003 & 0.004  & 249  & 1.07 \\
68.68101  & -16.6140  & 0.000 & 0.004  & 242  & 1.07 \\
68.75177  & -16.6223  & -0.005 & 0.004  & 213  & 1.35 \\
68.75423  & -16.6249  & -0.007 & 0.004  & 204  & 1.37 \\
68.78550  & -16.6218  & -0.009 & 0.004  & 155  & 1.66 \\
68.79312  & -16.6205  & -0.005 & 0.004  & 167  & 1.76 \\
69.58463  & -16.6143  & -0.002 & 0.003  & 332  & 1.06 \\
69.58902  & -16.6090  & 0.001 & 0.003  & 274  & 1.05 \\
69.66692  & -16.6107  & 0.001 & 0.003  & 286  & 1.05 \\
69.66967  & -16.6143  & 0.000 & 0.003  & 283  & 1.06 \\
70.60523  & -16.6138  & 0.002 & 0.003  & 337  & 1.03 \\
70.65614  & -16.6112  & -0.003 & 0.003  & 324  & 1.04 \\
70.73446  & -16.6386  & -0.009 & 0.003  & 259  & 1.28 \\
71.66974  & -16.6206  & -0.002 & 0.003  & 303  & 1.06 \\
71.67324  & -16.6252  & -0.002 & 0.003  & 307  & 1.07 \\
71.71607  & -16.6355  & -0.009 & 0.003  & 278  & 1.20 \\
\hline
\end{tabular}
\end{table}


\section{Early science results}
\begin{figure}
	\includegraphics[width=1.1\columnwidth, angle=0]{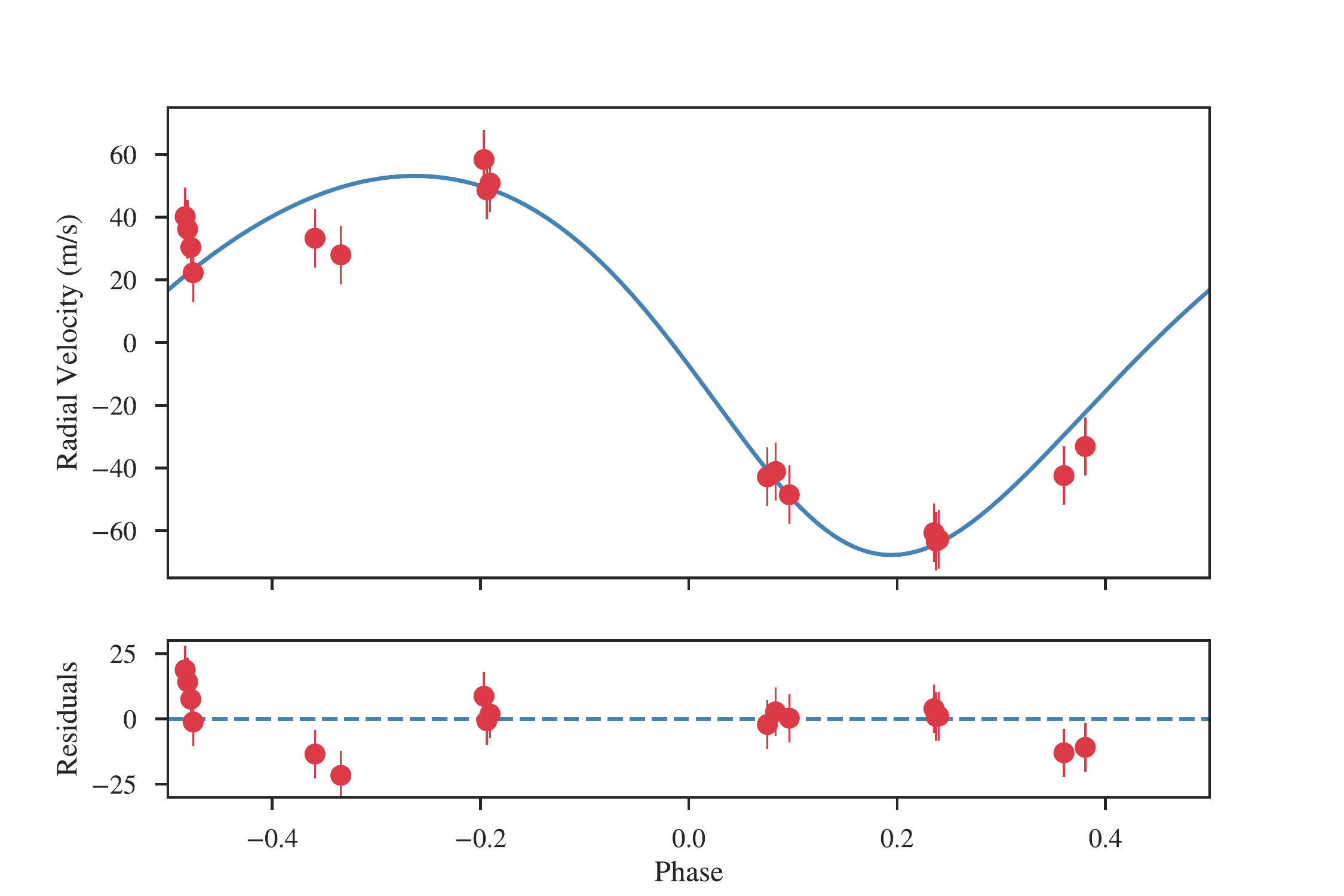}
    \caption{Detection of the known hot-jupiter orbiting the star HD 75289.}
    \label{planet}
\end{figure}

The main scientific motivation of FIDEOS is the detection and confirmation of exoplanets
by measuring precision RVs. To prove the scientific potential of the instrument we observed the known exoplanet system HD75289, which corresponds to a non-transiting hot Jupiter
(P=3.5 d, Mpsini = 0.47 M$_J$) orbiting a bright (V=6.4) G0V star \citep{Udry2000}. We observed this target with the ESO1.0m/FIDEOS system
for 6 consecutive nights obtaining between 2 and 3 spectra for each night. We adopted an exposure time of 600 and 900 seconds. Observations
were performed using the simultaneous calibration mode for airmass values between 1 and 1.4. The achieved SNR ranged between
130 and 225 per resolution element in the central part of the echelle order centred at 515 nm. Table \ref{tab-planet} presents
the measured RVs for HD75289 along with the bisector span measurements and other informative values.
The RVs are also presented in Fig \ref{planet} where a clear RV variation is identified. The amplitude of the variation is
consistent with the influence of a planetary companion. We used the radvel package \citep{Fulton2018} to find the orbital parameters.
We fixed the period to the known value of 3.509267 $\pm$ 0.000064 and considered as free parameters: the velocity semi-amplitude (K), the eccentricity, the pericenter argument, the time of pericenter passage (T0) and jitter.
We obtained K=60.0 +3.8/-3.6 ms$^{-1}$, eccentricity 0.136 +0.08/-0-07, pericenter argument 151 +18/-23, TO=2450829.91 $\pm$ 0.16 and jitter 2.4 +5.0/-2.1. All these values are consistent with \citep{Wang2011}. This result is a proof that the ESO1.0m/FIDEOS system can discover new extrasolar planets.

\begin{table}
\centering
\caption{Observations of HD75289. The columns are, BJD epoch, measured RV in km/s, Bisector span, exposure time, signal-to-noise ratio and airmass}
\label{tab-planet}
\begin{tabular}{lllllll} 
\hline
BJD-2458000 &  RV  & BS  &  BSe  &  t (exp)  &    SNR  &  AM   \\
\hline
66.84548 & 9.2742 & 0.036 & 0.005 & 600.0  & 138  & 1.13 \\
66.85294 & 9.2715 & 0.030 & 0.005 & 600.0  & 135  & 1.11 \\
66.86167 & 9.2721 & 0.029 & 0.005 & 600.0  & 131  & 1.09 \\
67.83206 & 9.3750 & 0.049 & 0.005 & 600.0  & 146  & 1.16 \\
67.84042 & 9.3710 & 0.055 & 0.005 & 600.0  & 148  & 1.13 \\
67.85111 & 9.3652 & 0.044 & 0.004 & 600.0  & 157  & 1.11 \\
67.85909 & 9.3571 & 0.047 & 0.004 & 600.0  & 157  & 1.09 \\
68.83879 & 9.3932 & 0.042 & 0.004 & 600.0  & 158  & 1.13 \\
68.84841 & 9.3835 & 0.040 & 0.004 & 900.0  & 204  & 1.11 \\
68.85930 & 9.3857 & 0.044 & 0.004 & 900.0  & 206  & 1.09 \\
69.79376 & 9.2920 & 0.045 & 0.004 & 600.0  & 179  & 1.29 \\
69.82118 & 9.2937 & 0.044 & 0.004 & 600.0  & 184  & 1.18 \\
69.86738 & 9.2863 & 0.044 & 0.004 & 900.0  & 218  & 1.07 \\
70.79275 & 9.2924 & 0.034 & 0.004 & 900.0  & 225  & 1.29 \\
70.86467 & 9.3017 & 0.054 & 0.004 & 900.0  & 172  & 1.07 \\
71.77872 & 9.3681 & 0.041 & 0.004 & 900.0  & 208  & 1.35 \\
71.86588 & 9.3628 & 0.033 & 0.004 & 900.0  & 196  & 1.06 \\
\hline
\end{tabular}
\end{table}

\section{Conclusions}
We presented results from the commissioning and early science programs of the spectrograph FIDEOS installed at the ESO 1m telescope of La Silla. The main results of our work are:

\begin{itemize}
\item[-] A new high resolution spectrograph optimised for precision RV measurements was successfully built at the Centre of Astro Engineering UC - AIUC. The new instrument is called FIber Dual Echelle Optical Spectrograph or FIDEOS.
\item[-] Our team refurbished the ESO 1m telescope of La Silla and equipped it with a new control system. 
\item[-] The new system ESO1.0m/FIDEOS saw first light in June 2016.
\item[-] The performances of FIDEOS meet the requirements and in particular we can reach a limiting magnitude of V=11 with S/N=10 in 30 min exposure time and a RV precision of $\sim$ 8 m/s on sky.
\item[-] We tested this system on a real science case and detected the known hot-jupiter HD 75289, obtaining parameters fully consistent with previous determinations.
\item[-] FIDEOS is accessible to the national and international community through collaboration with the AIUC and UCN, 10\% of the observing time will be available to the Chilean community.
\end{itemize}

\section*{Acknowledgements}
The FIDEOS project was funded by CONICYT through program FONDEF IdeA n. CA13I10203. Support was also provided by projects CONICY Anillo ACT-86, Anillo ACT-1417, Fondecyt n. 1130849, Fondecyt n. 1171364, and ESO Comite Mixto. 
MF was supported by project CONICY Gemini n. 32030014. AZ was supported by CONICYT grant n. 21170536. AJ acknowledges support from FONDECYT project n. 1171208 and from BASAL CATA PFB-06. AJ and RB  received support from project IC120009 ``Millennium Institute of Astrophysics (MAS)'' of the Millennium Science Initiative, Chilean Ministry of Economy. 

We are specially grateful to the administrative personnel of the AIUC Lilena Montenegro and Carlos Caire for their precious support in the management and logistic of the project. We also thank Christian Moni-Bidin at UCN, Juan Veliz at PUC, Jos\'{e} Pizarro at Universidad de Chile, the Department of Mechanics and Metallurgy of PUC and the personnel at ESO La Silla for their constant help during the execution of this project.

Finally a special thanks to Carlos Viscasillas, who visited the AIUC team during the early phases of the project, for his lasting interest in our work, his support and enthusiasm even from long distance.












\bsp	
\label{lastpage}
\end{document}